\documentclass[12pt,a4paper]{article}
\usepackage[utf8]{inputenc}
\usepackage{scalerel,stackengine,amsmath}
\usepackage{amsfonts}
\usepackage{amssymb}
\usepackage{hyphenat}
\usepackage{enumitem}
\usepackage{dsfont}
\usepackage[left=2cm,right=2cm,top=2.5cm,bottom=2.5cm]{geometry}
\usepackage{cite}
\usepackage{slashed}
\usepackage{hyperref}
\usepackage{graphicx}
\usepackage{caption}
\usepackage{float}

\usepackage{array}
\usepackage{physics}
\usepackage{float}
\usepackage{mathtools}
\usepackage{rotating}
\usepackage{amssymb}
\newcommand{\citet}{\cite} 
\usepackage{braket}
\usepackage[export]{adjustbox}
\usepackage{mathtools}
\usepackage{comment}
\usepackage{mathrsfs}  

\usepackage{adjustbox}
\usepackage{graphicx}
\usepackage{slashed}
\usepackage{mathrsfs}

\numberwithin{equation}{section}
\usepackage{lmodern}
\usepackage{bbold}
\usepackage{mmacells}
\usepackage[symbol]{footmisc}

\mmaDefineMathReplacement[≤]{<=}{\leq}
\mmaDefineMathReplacement[≥]{>=}{\geq}
\mmaDefineMathReplacement[≠]{!=}{\neq}
\mmaDefineMathReplacement[→]{->}{\to}[2]
\mmaDefineMathReplacement[⧴]{:>}{:\hspace{-.2em}\to}[2]
\mmaDefineMathReplacement{∉}{\notin}
\mmaDefineMathReplacement{∞}{\infty}

\mmaSet{
  morefv={gobble=2},
  linklocaluri=mma/symbol/definition:#1,
  morecellgraphics={yoffset=1.9ex}
}

\renewcommand{\a}{\alpha}
\renewcommand{\b}{\beta}
\renewcommand{\c}{\gamma}
\renewcommand{\d}{\delta}

\newcommand{\st}{{\fontfamily{cmss}\selectfont\text{sT}}}
\newcommand{\tran}{{\fontfamily{cmss}\selectfont\text{T}}}

\begin{document}

\begin{flushright}
July 2023
\end{flushright}

\bigskip

\begin{center}
{\bf \LARGE{\hspace*{-3mm}Minimal${}$\nolinebreak[4] Supergeometric${}$\nolinebreak[4] 
Quantum${}$\nolinebreak[4] 
Field\nolinebreak[4] 
Theories${}$} }
\end{center}

\vspace{5mm}

\begin{center}
{\large Viola Gattus\footnote[1]{E-mail address: {\tt
      viola.gattus@manchester.ac.uk}} and Apostolos
    Pilaftsis\footnote[2]{E-mail address: {\tt 
apostolos.pilaftsis@manchester.ac.uk}}}\\[3mm] 
{\it Department of Physics
and Astronomy, University of
Manchester,\\ Manchester M13 9PL, United Kingdom}
\end{center}
\vspace{2cm}

\centerline{\bf ABSTRACT}
\vspace{2mm}

\noindent
We formulate minimal SuperGeometric Quantum Field Theories (SG-QFTs) that allow for scalar-fermion field transformations in a manifestly reparameterisation covariant manner. First, we discuss the issue of uniqueness in defining the field-space supermetric of the underlying super\-manifold, and clarify the fact that different supermetric definitions can lead to distinct theories in the off-shell kinematic region. By adopting natural choices for the field-space supermetric, we~then show that scalar fields alone cannot induce a non-trivial field-space Riemannian curvature in the fermionic sector, beyond the one originating from the scalar part of the theory. We~present for the first time minimal SG-QFT models that feature non-zero fermionic curvature both in two and four spacetime dimensions. Physical applications of SG-QFTs are discussed.


\thispagestyle{empty}

\newpage

\section{Introduction}

For more than half a century, covariant and differential geometric methods still continue to play a central role in the development of Quantum Field Theory (QFT)~\cite{DeWitt:1967ub}. 
Besides aspects of gauge covariance in effective actions~\cite{Gaillard:1985uh,Pilaftsis:1996fh,Cornwall:2010upa,Binosi:2009qm}, these methods were used to compute transition amplitudes of chiral loops in a reparameterisation invariant manner~\cite{Honerkamp:1971sh,Ecker:1972tii}. They have also been applied within the context of non-linearly realised supersymmetric theories~\cite{Alvarez-Gaume:1981exa}. Beyond the classical approximation, these covariant and differential geometric methods have been put on a more rigorous footing by Vilkovisky and DeWitt~(VDW)~\cite{Vilkovisky:1984st,DeWitt:1985sg} to address the issue of gauge-fixing parameter independence in gauge and quantum gravity theories. This VDW framework was developed further by several other authors~\cite{Barvinsky:1985an,Ellicott:1987ir,Burgess:1987zi,Odintsov:1989gz}. More recently, a related differential geometric formalism was utilised to resolve the so-called quantum frame problem in cosmological single-field and multi-field inflation~\cite{Kamenshchik:2014waa,Burns:2016ric,Karamitsos:2017elm}, along with the issue of uniqueness of the path-integral measure of the VDW effective action~\cite{Finn:2019aip,Falls:2018olk} beyond the Born approximation. On the other hand, similar geometric techniques were employed to analyse new-physics phenomena within the framework of Effective Field Theories (EFTs) beyond the Standard Model~(SM)~\cite{Alonso:2016oah,Nagai:2019tgi,Helset:2020yio,Cohen:2021ucp,Talbert:2022unj,Helset:2022tlf}, also known as SMEFT (for a recent review, see~\cite{Isidori:2023pyp}).
 
In spite of the enormous progress made in field-theoretic differential geometric techniques for bosonic theories, the inclusion of fermions in the VDW formalism encountered a number of theoretical difficulties and limitations. Specifically, earlier attempts to include fermions in a reparameterisation and frame-invariant manner were either related to the geometry of bosons through supergravity~\cite{Alvarez-Gaume:1981exa} or dependent on a specific EFT-operator~\cite{Helset:2020yio,Fumagalli:2020ody,Talbert:2022unj}. Unlike the commuting boson fields, fermion fields are anti-commuting Grassmannian variables, so considering them consistently as independent chart variables in the path-integral configuration space requires the consideration of differential {\em supergeometry} (SG) on {\em super\-manifolds}~\cite{DeWitt:2012mdz}. 

Recently, a manifestly reparameterisation-invariant formulation of scalar-fermion theories was put forward~\cite{Finn:2020nvn}, where the field-space metric was defined from the action. The formalism enabled to obtain earlier known results of the effective action at the one-loop level, but also a new expression for the SG effective action at the two-loop order. However, definite models with non-zero fermionic curvature have not been presented yet in the existing literature. 

In this paper we present for the first time novel minimal SG-QFT models that feature non-zero fermionic curvature both in two and four spacetime dimensions. In formulating these minimal models, we discuss the issue of uniqueness in defining from the action the field-space metric of the underlying supermanifold, ${}_\alpha G_\beta$, which is also termed  {\em supermetric}. We explain how different definitions of the supermetric ${}_\alpha G_\beta$ will usually lead to distinct theories in the off-shell kinematic region. However, some natural choices can be made that rely on a model-function which appears in the kinetic term of the fermions. As noted in~\cite{Finn:2020nvn}, the derived metric of the field-space supermanifold should be {\em supersymmetric}, which means that~${}_\alpha G_\beta$ should be invariant under the operation of supertransposition (\text{\sf sT}) to be defined in Section 2. Moreover, we show that scalar fields alone cannot induce a non-trivial field-space Riemannian curvature in
the fermionic sector, beyond the one that has its origin in the scalar part of the theory. 

The paper is organised as follows. After this brief introductory section, we discuss in Section~\ref{sec:SG} the basic covariant structure of scalar-fermion SG-QFTs including their key model functions, ${}_\alpha k_\beta$ and $\zeta^\mu_\alpha$. Here we ignore the effect of gravity and gauge interactions, and leave their study in future works. In the same section, we also present our approach to deriving the supermetric from the classical action of an SG-QFT. In Section~\ref{sec:Models} we first show a no-go theorem for the generation of a non-zero super-Riemannian curvature in a bilinear kinetic fermionic sector from the existence of scalar fields only in SG-QFTs. Then, we present two minimal models that realise non-zero fermionic curvature when the model function $\zeta^\mu_\alpha$ contains non-linear fermionic terms.
Section~\ref{sec:Superprops} presents the scalar-fermion superpropagators, as well as the three- and four-point supervertices, thereby generalising earlier results that were derived in pure bosonic theories. Finally, Section~\ref{sec:Summary} summarises the main results of our study and provides an outlook for physical applications within this novel SG-QFT framework.

\section{Supergeometry on the Scalar-Fermion Field Space}\label{sec:SG}

Let us briefly review some basic aspects of differential supergeometry on the scalar-fermion field space~\cite{DeWitt:2012mdz} that are relevant to the formulation of SG-QFTs. To start with, we note that a set of $N$ real scalar fields and $M$ Dirac fermions describe a field-space supermanifold of dimension $(N|8M)$ in four spacetime dimensions (4D). A chart of this supermanifold may be denoted by
\begin{equation}
   \label{eq:PhiChart}
  \boldsymbol{\Phi}\ \equiv\ \left\{\Phi^{\alpha}\right\}\ =\ \left(\phi^{A}\:,\: \psi^{X}\:,\: \overline{\psi}{}^{Y,\tran}\:\right)\!{}^\tran \,.
\end{equation}
Here and in what follows, a Greek index  like $\a=1,2,\dots, N+M$ labels all fields. Otherwise, we will use Latin letters from the beginning of the alphabet to denote individual bosonic degrees of freedom and letters from the end to denote fermionic ones. In~analogy\- to the standard theory of manifolds, general field reparameterisations of the form,
\begin{equation}
   \label{eq:PhiDiff}
    \Phi^{\alpha}\ \rightarrow\ \widetilde{\Phi}^{\alpha}\, =\, \widetilde{\Phi}^{\alpha}(\boldsymbol{\Phi})\;,
\end{equation}
become now diffeomorphisms on the supermanifold. Notice that the class of transformations in~\eqref{eq:PhiDiff} cover any {\em ultralocal} redefinitions of the scalar and fermions fields without  introducing extra spacetime derivatives of fields like $\partial_\mu \Phi^{\alpha}$, along the lines of the VDW formalism. 

Up to second order in $\partial_{\mu} \Phi^{\alpha}$, the Lagrangian for a general scalar-fermion theory, which is invariant under field-space diffeomorphisms,  can be written in terms of three model functions: (i)~a rank-2 field-space tensor $_{\alpha} k_{\beta}(\boldsymbol{\Phi})$, (ii)~a mixed spacetime and field-space vector $\zeta_{\alpha}^{\mu}(\boldsymbol{\Phi})$, and (iii)~a zero-grading scalar~$U(\boldsymbol{\Phi})$ describing the potential and Yukawa interactions. Such a diffeomorphically- or frame-invariant Lagrangian reads \cite{Finn:2020nvn}
\begin{equation}
   \label{eq:LSGQFT}
    \mathcal{L}\ =\ \frac{1}{2} g^{\mu \nu} \partial_{\mu} \Phi^{\alpha}\:_{\alpha} k_{\beta}(\boldsymbol{\Phi})\: \partial_{\nu} \Phi^{\beta}\: +\: \frac{i}{2}\, \zeta_{\alpha}^{\mu}(\boldsymbol{\Phi})\: \partial_{\mu} \Phi^{\alpha}\: -\: U(\boldsymbol{\Phi}).
\end{equation}
In~\eqref{eq:LSGQFT}, $_{\alpha} k_{\beta}$
vanishes when the indices $\alpha$ or $\beta$ are fermionic, i.e.~${}_X k_A = {}_A k_Y = {}_X k_Y = 0$. Note that $_{\alpha} k_{\beta}$~plays the role of the field-space metric \cite{Finn:2019aip} for a purely bosonic theory. Therefore, the function $\zeta_{\alpha}^{\mu}$ is introduced to describe the fermionic sector. Notice that $\zeta_{\alpha}^{\mu}$ may also be used to include chiral fermions by decomposing each Dirac fermion into pairs of Majorana fermions.

The model functions, $_{\alpha} k_{\beta}$ and $\zeta_{\alpha}^{\mu}$, can unambiguously be extracted from the Lagrangian according to the following prescription~\cite{Finn:2020nvn}:
\begin{equation}
   \label{eq:kzeta}
{ }_{\alpha} k_{\beta}\ =\ \frac{g_{\mu \nu}}{D} \frac{\overrightarrow{\partial}}{\partial\left(\partial_{\mu} \Phi^{\alpha}\right)} \mathcal{L} \frac{\overleftarrow{\partial}}{\partial\left(\partial_{\nu} \Phi^{\beta}\right)}\:, \qquad 
\zeta_{\alpha}^{\mu} =\frac{2}{i}\left(\mathcal{L}-\frac{1}{2} g^{\mu \nu} \partial_{\mu} \Phi^{\gamma}{}_\gamma k_{\delta}\, \partial_{\nu} \Phi^{\delta}\right) \frac{\overleftarrow{\partial}}{\partial\left(\partial_{\mu} \Phi^{\alpha}\right)}\; .
\end{equation}
To equip the supermanifold with a metric, we need to construct a pure field-space covector~$\zeta_\alpha$ from $\zeta_{\alpha}^{\mu}$. As it is evident from~\eqref{eq:LSGQFT} and explained in~\cite{Finn:2020nvn}, the Lorentz index $\mu$ in $\zeta_{\alpha}^{\mu}$ can only arise from the presence of a $\gamma^\mu$-matrix, or a $\sigma^\mu = (\sigma^0\,,\boldsymbol{\sigma})$ matrix in the chiral basis, where~$\boldsymbol{\sigma} = (\sigma^1\,, \sigma^2\,, \sigma^3)$ are the three Pauli matrices.

To find the metric of the field-space supermanifold, it proves useful to distinguish two categories of SG-QFTs depending on the actual structure of the model function~$\zeta_{\alpha}^{\mu}$. In the first category, the fermionic components of $\zeta_{\alpha}^{\mu}$ may be expressed in a factorisable form as
\begin{equation}
  \label{eq:zetagamma}
    \zeta^{\mu}_{\:\a}=\zeta_{\:\b}\: ^{\b}(\Gamma^{\mu})_{\:\a},\qquad \text{where}\quad \Gamma^{\mu}  =\left(\begin{array}{cc}
\gamma^{\mu} & 0 \\
0 & \left(\gamma^{\mu}\right)^{\tran}
\end{array}\right).
\end{equation}
The second category of SG-QFTs does not possess the factorisation property~\eqref{eq:zetagamma}. As we will see in Section~\ref{sec:Models}, the distinction between factorisable and non-factorisable $\zeta_{\alpha}^{\mu}$ affects the geometric properties of the field-space supermanifold. For the first category of SG-QFTs, it is straightforward to project a proper field-space covector $\zeta_{\:\a}$ from $\zeta_{\alpha}^{\mu}$ given in \eqref{eq:zetagamma}. The simplest way would be to introduce a differentiation with respect to the $\gamma^\mu$ matrix as done in~\cite{Finn:2020nvn}, i.e. 
\begin{equation}
   \label{eq:zeta}
    \zeta_{\alpha}\: =\: \frac{1}{D}\, \frac{\delta \zeta_{\alpha}^{\mu}}{\delta \gamma^{\mu}}\; ,
\end{equation}
where $D$ is the number of space-time dimensions. But for the second category of SG-QFTs for which $\zeta^{\mu}_{\:\a}$ does not obey~\eqref{eq:zetagamma}, one may alternatively use the more natural projection operation,
\begin{equation}
   \label{eq:projzeta}
     \zeta_{\beta}^{\mu} \:\:^\b\left(\overleftarrow{\Sigma}_{\mu}\right)_{\alpha}\: =\: \zeta_{\alpha}\;, \qquad \text{where}\quad 
  \overleftarrow{\Sigma}_{\mu}  =\frac{1}{D}\left(\begin{array}{cc}
\frac{\overleftarrow{\partial}}{\partial \gamma^{\mu}} & 0 \\
0 & \Gamma_{\mu}
\end{array}\right)\;.
\end{equation}
In the above, the differentiation acting on the fermionic components of $\zeta_{\alpha}^{\mu}$ is replaced by contraction with $\gamma^{\mu}$ matrices. In this way, the spin-$3/2$ degrees of freedom (dofs) contained in~$\zeta^{\mu}_{a}$ are projected onto spin-$1/2$ dofs in $\zeta_a$. In this study we adopt the projection method~\eqref{eq:projzeta} which can be applied to both categories of SG-QFTs.  

An important geometric property of SG-QFTs as described by the Lagrangian~$\mathcal{L}$ in \eqref{eq:LSGQFT} is that $\mathcal{L}$ is a scalar in the field-space supermanifold.  In other words, $\mathcal{L}$ remains invariant under the field redefinitions in~\eqref{eq:PhiDiff}, provided all model functions and the field-space tangent vectors are appropriately transformed. In this SG framework, we have 
\begin{equation}
    \partial_{\mu}\widetilde{\Phi}^\a(\boldsymbol{\Phi})\: =\: \partial_{\mu}\Phi^\b(\boldsymbol{\Phi}) \:_{\b }J^\a(\boldsymbol{\Phi})\;,
\end{equation}
where $\:_{\a }J^\b=\:_{\a,}\widetilde{\Phi}^\b$ is the Jacobian of the transformation and the subscript $\alpha$ before the comma on the left side of $\widetilde{\Phi}^\b$ denotes ordinary left-to-right differentiation with respect to the field $\Phi^\a$.

A field-space supermanifold of interest to us must be endowed with a rank-2 field-space tensor~$_\a G_\b$, which is supersymmetric, i.e. 
\begin{equation}
    {}_\alpha G_\beta\: =\: ({}_\alpha G_\beta)^{\sf sT}\: =\: (-1)^{\alpha +\beta +\alpha\beta} {}_\beta G_\alpha\;,
\end{equation}
and non-singular. Such a supermanifold is called Riemannian~\cite{DeWitt:2012mdz} and the rank-2 field-space tensor~$_\a G_\b$ is known as the \textit{supermetric}. Its inverse $^\a G^\b$, deduced from the identity: $ _\a G_\c \:G^{\c \b}=\: _\a\delta^{\b}$, satisfies
\begin{equation}
    ^\a G^\b\: =\: G^{\a \b}\: =\: (-1)^{\a \b}\: G^{\b \a} .
\end{equation}
In the above, we have employed the compact index calculus and conventions by DeWitt in~\cite{DeWitt:2012mdz}, so that the exponents of $(-1)$ determine the grading of the respective quantities and take the values~$0$ or~$1$ for commuting or anticommuting fields, respectively. According to  DeWitt's conventions, the usual tensor contraction between indices can only be performed if the two indices to be summed over are adjacent. Otherwise, extra factors of $(-1)$ must be introduced whenever two indices are swapped.

Given the supermetric ${}_\alpha G_\beta$, the Christoffel symbols
$\Gamma^{\a}_{\:\:\b \c}$ can be evaluated as in~\cite{DeWitt:2012mdz}, from which the super-Riemann tensor is obtained
\begin{equation}
   \label{eq:Riemann}
R^\a_{\:\:\b \c \delta}\: =\: -\,\Gamma^{\a}_{\:\:\b \c,\delta}\: +\: (-1)^{\c \delta}\: \Gamma^{\a}_{\:\:\b \delta,\c}\: +\:(-1)^{\c(\sigma+\b)}\: \Gamma_{\:\:\sigma \c}^{\a} \Gamma_{\:\:\b \delta}^{\sigma}\: -\: (-1)^{\delta(\sigma+\b+\c)}\: \Gamma_{\:\:\sigma \delta}^{\a} \Gamma_{\:\:\b \c}^{\sigma}\; .
\end{equation}
The super-Ricci tensor is obtained by contracting the first and third indices of the super-Riemann tensor~\cite{DeWitt:2012mdz},
\begin{equation}
    R_{\a \b}\: =\: (-1)^{\c (\a + 1)}R^{\c}_{\:\:\a \c \b}\; .
\end{equation}
 Further contraction of the remaining two indices of $R_{\a \b}$ yields the super-Ricci scalar,
\begin{equation}
  \label{eq:Rscalar}
    R\: =\:  R_{\a \b}\:G^{\b \a}\; .
\end{equation}
Note that the super-Ricci tensor is supersymmetric, i.e.~$R_{\a \b} = (R_{\a \b})^{\sf sT} =
(-1)^{\a \b}R_{\b \a}$.

To determine the supermetric $_\a G_\b$ of the scalar-fermion field space, we follow the procedure presented in~\cite{Finn:2020nvn, Pilaftsis:2022las}. After calculating the projected model function~$\zeta_\a$ as stated in~\eqref{eq:projzeta}, we may now construct the rank-2 field-space anti-supersymmetric tensor
\begin{equation}
  \label{eq:lambda}
   {}_{\alpha} \lambda_{\beta}\ =\ \frac{1}{2}\, \Big({}_{\a,}\zeta_{\beta}\: -\: (-1)^{\alpha+\beta+\alpha \beta}  \:_{\b,}\zeta_{\alpha}\Big)\; .
\end{equation}
Exactly as happens for the anti-symmetric field strength tensor $F_{\mu \nu}$ in QED in curved spacetime, the derivatives appearing in \eqref{eq:lambda} are ordinary derivatives and not covariant ones, since the Christoffel symbols drop out for such constructions of anti-supersymmetric rank-2 tensors like~$_\a \lambda_\b$.

The so-constructed ${ }_{\alpha} \lambda_{\beta}$ turns out to be singular in the presence of scalar fields, and so the scalar contribution $_{\alpha} k_{\beta}$  has to be added which results in the new rank-2 field-space tensor,
\begin{equation}
   \label{eq:Lambda}
_{\alpha} \Lambda_{\beta}\ =\:  _{\alpha} k_{\beta} \:+\:  _{\alpha} \lambda_{\beta}\; .
\end{equation}
However, $_{\alpha} \Lambda_{\beta}$ cannot act as a supermetric, since it
is not supersymmetric. To find a suitable rank-2 tensor that satisfies the latter property, we make use of the vielbein formalism~\cite{Schwinger:1963re, Yepez:2011bw} which allows to
compute the field-space vielbeins $_\a e^a$, if the form of $_{\alpha} \Lambda_{\beta}$ is known in the local field-space frame. For the latter, we demand that the Lagrangian~\eqref{eq:LSGQFT} assumes the canonical Euclidean form in this local frame. In this way, we may compute the field-space supermetric as~\cite{Finn:2020nvn}
\begin{equation}
  \label{eq:aGb}
    _\a G _\b \:=\: _\a e^a\:\:_a H_b\:\:^be^{\st}_\b\;,
\end{equation}
where 
\begin{equation}    
   \label{eq:aHb}
{ }_{a} H_{b} \equiv\left(\begin{array}{ccc}
{\bf 1}_{N} & 0 & 0 \\
0 & 0 & {\bf 1}_{4M}  \\
0 & -{\bf 1}_{4M} & 0 
\end{array}\right)
\end{equation}
is the local field-space metric in 4D.

Finally, according to VDW formalism~\cite{Vilkovisky:1984st,DeWitt:1985sg}, we need to promote the field space to a configuration space, so as to take into account the spacetime dependence of the fields. In this configuration space, the coordinate charts are extended as
\begin{equation}
  \label{eq:PhiCS}
    \Phi^{\hat{\alpha}}\: \equiv\: \Phi^{\alpha}\left(x_{\alpha}\right),
\end{equation}
where $x_\a$ is the spacetime coordinate of a generic field $\Phi^\a$. Likewise,
the supermetric gets generalised as 
\begin{equation}
   \label{eq:CSGmetric}
    { }_{\hat{\alpha}} G_{\hat{\beta}}\ =\ { }_{\alpha} G_{\beta}\: \delta(x_{\alpha} - x_{\beta})\;,
\end{equation}
where $\delta(x_{\alpha} - x_{\beta})$ is the $D$-dimensional $\delta$-function.
This generalisation affects the Christoffel symbols and the Riemann tensors, as given in more detail in~\cite{Finn:2019aip}.

\section{Minimal Models}\label{sec:Models}

In this section we first show a no-go theorem that no non-zero field-space curvature can be generated in the fermionic sector from a single scalar field and multiple fermion species, as long as the model function $\zeta^\mu_\alpha$ only contains linear terms in the fermionic fields. We then present two minimal SG-QFT models with non-zero fermionic curvature which is induced by the addition of non-linear terms in fermion fields to $\zeta^\mu_\alpha$.

\subsection{No-Go Theorem on Fermionic Field-Space Curvature}

The simplest case with one bosonic field $\phi$ and one Dirac fermion $\psi$ was analysed in~\cite{Finn:2020nvn}. This case reduces to a flat field space, and as such, we will not repeat it here.
Instead, we consider a more general scenario with a single boson $\phi$ and a multiplet
$\boldsymbol{\psi} =\{ \psi^X\} $ of Dirac fermion fields  (with $X=1,2,\dots M$). In 4D, such a scenario has $(1|8M)$ field-space coordinates. Up to second order in spacetime derivatives, the Lagrangian for such a system is given  by
\begin{equation}
   \label{eq:NoGo}
\begin{aligned}
\mathcal{L}\ =&\ \frac{1}{2}\,k(\phi)\, (\partial^{\mu} \phi)\, (\partial_{\mu} \phi) \: -\: \frac{1}{2} h_{X Y}(\phi)\, \overline{\psi}^{X} \gamma^{\mu} \psi^{Y} (\partial_{\mu} \phi) \\
&+\: \frac{i}{2}\, g_{X Y}(\phi)\left[\overline{\psi}^{X} \gamma^{\mu} (\partial_{\mu} \psi^{Y})\, -\, (\partial_{\mu} \overline{\psi}^{X}) \gamma^{\mu} \psi^{Y}\right]\;. 
\end{aligned}
\end{equation}
Evidently, the single field $\phi$ cannot induce by itself a non-zero Riemannian curvature in the scalar sector. Hence, if there is a non-trivial field-space curvature, this can only come from the fermionic sector of the Lagrangian~\eqref{eq:NoGo}.

If the field space is flat, one should be able to find a suitable reparameterisation of the fields to bring the Lagrangian~\eqref{eq:NoGo} into a canonical Cartesian form. To this end, let us consider the following redefinition of fermionic fields:
\begin{equation}
   \label{eq:psitilde}
    \boldsymbol{\psi}\ \longrightarrow\ \widetilde{ \boldsymbol{\psi}}\, =\, \boldsymbol{K}(\phi)^{-1}\,\boldsymbol{\psi}\;,
\end{equation}
where $\boldsymbol{K}$ is a $4M\times 4M$-dimensional matrix that 
only depends on the scalar field~$\phi$. 
The field reparameterisation~\eqref{eq:psitilde} modifies the fermionic part of the Lagrangian \eqref{eq:NoGo} as follows:
\begin{equation}
   \label{eq:NoGoLf}
    \begin{aligned}
\mathcal{L}_{\text{f}}\ =&\ 
-\frac{1}{2}\left[h_{X Y} K_{W}^{\dagger X} K_{Z}^{Y}-i g_{X Y} 
\left( K_{W}^{\dagger X} \frac{\partial K_{Z}^{Y}}{\partial \phi}-\frac{\partial K_{W}^{\dagger X}}{\partial \phi} K_{Z}^{Y}\right)\right] \widetilde{\overline{\psi}}{}^{W} \gamma^{\mu} \widetilde{\psi}^{Z}\,(\partial_{\mu} \phi)\\
&\ +\frac{i}{2} g_{X Y} K_{W}^{\dagger X} K_{Z}^{Y}\left[\widetilde{\overline{\psi}}{}^{W} \gamma^{\mu} (\partial_{\mu} \widetilde{\psi}^{Z})\, -\, (\partial_{\mu} \widetilde{\overline{\psi}}{}^{W}) \gamma^{\mu} \widetilde{\psi}^{Z}\right]\; .
\end{aligned}
\end{equation}
To put Lagrangian~\eqref{eq:NoGo} into a canonical form, it suffices to eliminate the first term on the RHS of~\eqref{eq:NoGoLf}. 
We therefore require the vanishing of the matrix expression,
\begin{equation}
   \label{constraint}
    i\,\boldsymbol{K}^{\dagger}\boldsymbol{h} \boldsymbol{K}\: +\: \boldsymbol{K}^{\dagger} \boldsymbol{g}\frac{\partial \boldsymbol{K}}{\partial\phi}\: -\: \frac{\partial \boldsymbol{K}^\dagger}{\partial \phi}\boldsymbol{g}\boldsymbol{K}\ =\ {\bf 0}\; ,
\end{equation}
with $\boldsymbol{g} =\{ g_{XY}\}$ and $\boldsymbol{h} =\{ h_{XY}\}$.
Even though it is straightforward to find a solution to~\eqref{constraint} for the case of a single fermion field~\cite{Finn:2020nvn}, it becomes non-trivial in the presence of many fermions. However, after some familiarisation with the analytic expression on the LHS of~\eqref{constraint}, a simple and intuitive solution can be obtained, given by
\begin{equation}
    \boldsymbol{K}(\phi)\ =\ \exp\left(- \frac{i}{2}\int_{0}^{\phi}\boldsymbol{g}^{-1}\boldsymbol{h}\: d \phi\right)\; .
\end{equation}
Consequently, the field-space supermanifold for this theory is flat. 

It is worth noting here that one can confirm the field-space flatness of the theory within our formalism of an SG-QFT as well. In detail, the supermetric derived from~\eqref{eq:NoGo} reads~\cite{Finn:2020nvn, Pilaftsis:2022las}
\begin{equation}
   \label{eq:metric}
    { }_{\alpha} G_{\beta}\ =\ \left(\begin{array}{ccc}
k-\frac{1}{2}\boldsymbol{\overline{\psi}}\left(\boldsymbol{g}^{\prime}- i \boldsymbol{h}\right)\boldsymbol{g}^{-1}\left(\boldsymbol{g}^{\prime} + i \boldsymbol{h}\right)\boldsymbol{\psi} & -\frac{1}{2}\boldsymbol{\overline{\psi}}\left(\boldsymbol{g}^{\prime}- i \boldsymbol{h}\right) & \frac{1}{2}\boldsymbol{\psi}^{\:\tran}\left(\boldsymbol{g}^{\prime\: \tran}  + i \boldsymbol{h}^\tran\right)\\
\frac{1}{2}\left(\boldsymbol{g}^{\prime\:\tran}- i \boldsymbol{h}^\tran\right) \boldsymbol{\overline{\psi}}^{\:\tran} & 0 & \boldsymbol{g}^\tran 1_{4} \\
-\frac{1}{2}\left(\boldsymbol{g}^{\prime}+ i \boldsymbol{h}\right) \boldsymbol{\psi} & -\boldsymbol{g} 1_{4} & 0
\end{array}\right)\;.
\end{equation}
However, as opposed to what was conjectured in~\cite{Finn:2020nvn}, we find that the super-Riemann tensor computed from the field-space supermetric~\eqref{eq:metric} vanishes identically, thus implying that the field-space supermanifold is flat. This exercise shows that the scalar field $\phi$ acts only as an external parameter in the fermionic sector. If we add more scalar fields, this last result will not alter as long as the model function ${}_\alpha k_\beta \equiv k_{AB}$ does not introduce any curvature in the scalar sector, in which case $k_{AB}$ can then be brought into a canonical (Cartesian) form, i.e.~$k_{AB} = \delta_{AB}$. The rest of the proof goes along the lines discussed above for the single scalar case. In addition, one would need to consider an extra index $A = 1,2,\dots N$ counting the $N$ scalars~$\phi^A$. As a consequence, one would have to solve a set
of independent $N$ matrix equations like~\eqref{constraint}, resulting in $N$ different matrices $\boldsymbol{K}_A(\phi^A)$. This concludes our proof of the no-go theorem.

For a pure fermionic theory, non-zero fermionic curvature effects can only then be generated if the model function $\zeta^\mu_\alpha$ depends non-linearly on the fermion fields. In the next two subsections, we present two SG-QFT models that realise non-zero fermionic curvature.

\subsection{Non-zero Fermionic Field-Space Curvature: Model I}

Let us first consider a 2D SG-QFT model, also called Model I, which includes one scalar field~$\phi$ and one Dirac fermion represented as~$\psi^\tran = (\psi_1\,, \psi_2)$. The Lagrangian of this simple model is given by
\begin{equation}
   \label{eq:ModelI}
\begin{aligned}
\mathcal{L}_{\mathrm{I}}\ =&\ \frac{1}{2} k\, (\partial_{\mu}\phi)\, (\partial^{\mu} \phi)\: +\: \frac{i}{2} \left(g_0 +g_1\overline{\psi}\psi \right)\left[\overline{\psi} \gamma^{\mu} (\partial_{\mu} \psi)\,-\, (\partial_{\mu} \overline{\psi}) \gamma^{\mu} \psi\right]\: +\:
Y\,\overline{\psi} \psi\: -\: V\;,
\end{aligned}
\end{equation}
where $\gamma^\mu = (\sigma^1\,, -i\sigma^2)$. Here, all the model functions~$k$, $g_0$,  $g_1$, $Y$ and $V$ depend on the scalar field~$\phi$. Note that the model function $\zeta^\mu_\alpha$ derived from~\eqref{eq:ModelI} takes on the factorisable form of~\eqref{eq:zetagamma}, with
\begin{equation}
   \label{eq:zetaMI}
\zeta_\a\ =\ \Big\{0\,, \big(g_0 + g_1\overline{\psi}\psi\big)\,\overline{\psi}\,, \big(g_0 + g_1\overline{\psi}\psi\big)\,\psi^{\tran}\Big\}\;.
\end{equation}
Using the method of~\cite{Finn:2020nvn} briefly outlined in Section~2, we may derive the field-space super\-metric~$\boldsymbol{G} = \{ {}_\alpha G_\beta\}$ in the superspace $\boldsymbol{\Phi}^\tran = (\phi\,, \psi^{\tran}\,, \overline{\psi})$,
\begin{equation}\label{metric_2d_general}
    \boldsymbol{G}\, =\,\left(\begin{array}{crr}
k\: +\: b^\tran (d^{-1})^\tran a^\tran -\, a\,d^{-1}\,b\: 
& \ -a &\ b^\tran\\
\hspace{3mm} a^\tran &\ 0 &\ d^\tran\\
-b &\ -d & 0\hspace{2mm}{}
\end{array}\right),
\end{equation}
where
\begin{equation}
\begin{aligned}
a\ =\ \frac{1}{2}\,\overline{\psi}\left(g_0^\prime + g_1^\prime \overline{\psi}\psi\right)\;,\quad
b\ =\ \frac{1}{2}\left(g_0^\prime + g_1^\prime \overline{\psi}\psi\right)\psi\; ,\quad
d\ =\ \left(g_0+ g_1\overline{\psi}\psi\right) 1_2\, +\, g_1\psi\overline{\psi}\; ,
\end{aligned}
\end{equation}
and a prime $({}^\prime)$ on the model functions~$g_{0,1}$ denotes differentiation with respect to $\phi$. Note that $\boldsymbol{G}$ is supersymmetric, since $b^\tran (d^{-1})^\tran a^\tran = -a\,d^{-1} b$. 

Given the supermetric~$\boldsymbol{G}$, we may now compute the  
non-zero components of the Riemann tensor. For instance, if
$g_0=g_1=1$, these components are found to be
\begin{equation}
   \label{eq:RiemannMI}
\begin{aligned}
R^{\psi_1}_{\:\:\psi_1\overline{\psi}_1\psi_2}&=-R^{\overline{\psi}_2}_{\:\:\overline{\psi}_2\overline{\psi}_1\psi_2}= \psi_1\overline{\psi}_2-1\;, \\
 R^{\psi_1}_{\:\:\psi_1\overline{\psi}_2\psi_2}&=R^{\overline{\psi}_1}_{\:\:\overline{\psi}_2\overline{\psi}_1\psi_2}=-R^{\overline{\psi}_1}_{\:\:\overline{\psi}_1\overline{\psi}_2\psi_2}= -\psi_1\overline{\psi}_1\;, \\
 R^{\psi_2}_{\:\:\psi_1\overline{\psi}_1\psi_2}&= R^{\overline{\psi}_2}_{\:\:\overline{\psi}_2\overline{\psi}_1\psi_1}=-R^{\psi_2}_{\:\:\psi_2\overline{\psi}_1\psi_1} =\psi_2\overline{\psi}_2\;, \\
 R^{\psi_2}_{\:\:\psi_1\overline{\psi}_2\psi_2}&=R^{\overline{\psi}_1}_{\:\:\overline{\psi}_1\overline{\psi}_2\psi_1}=-R^{\psi_2}_{\:\:\psi_2\overline{\psi}_2\psi_1}=-R^{\overline{\psi}_1}_{\:\:\overline{\psi}_2\overline{\psi}_1\psi_2}= 1-\psi_2\overline{\psi}_1\;.
\end{aligned}
\end{equation}
Hence, the minimal SG-QFT model of~\eqref{eq:ModelI} exhibits a non-zero fermionic field-space curvature. Allowing for $\phi$-dependent model functions $g_{0,1}$, the super-Ricci scalar evaluates to\footnote[3]{In our study, we 
did not pay attention to the dimensionality of the model parameters, $k$, $g_0$, $g_1$, and their derivatives $g'_0$, $g'_1$, that enter the supermetric $\boldsymbol{G}$ from which $R$ is evaluated. Their dimensionality can be restored following the approach in~\cite{Finn:2019aip} to ensuring uniqueness of the path-integral measure. In detail, the energy~($E$) dimension of $R$ is the same as the inverse of the squared field-space line element: $d\Sigma^2 = d\Phi^\a{}_\a G_\b\, d\Phi^\b$, i.e.~$[R] = [d\Sigma^2]^{-1}$ in any $D$ dimensions. Like in~\cite{Finn:2019aip}, we take $d\Sigma^2$ to be dimensionless,
whilst considering the dimensions of the fields: $[\phi ] = E^{(D-2)/2}$, $[\psi ] = E^{(D-1)/2}$.
To do so, we rescale the model parameters $k$, $g_{0,1}$ as: $\tilde{k} = \ell^{D-2} k$, 
$\tilde{g}_{0,1} = \ell^{D-1} g_{0,1}$, where $\ell = \ell({\bf \Phi})$ plays the role of an effective Planck length, with $[\ell ] = E^{-1}$. In addition, $\tilde{g}_1$ should be divided by the energy cut-off $\Lambda^{D-1}$ that normalises
the fermionic bilinear $\overline{\psi}\psi$, in the context of an effective field theory. Hence, in $D$ dimensions we have: 
$[\tilde{k}] = E^{-(D-2)}$,
$[\tilde{g}_0] = E^{-(D-1)}$, $[\tilde{g}_1] = E^{-2(D-1)}$,
$[\tilde{g}'_0] = E^{2 -3D/2}$ and $[\tilde{g}'_1] = E^{3 -5D/2}$. One may then verify that with the rescaled parameters~$\tilde{k}$, $\tilde{g}_{0,1}$ and $\tilde{g}'_{0,1}$, the so-redefined super-Ricci scalar $\widetilde{R}$ is dimensionless in~\eqref{eq:RicciMI} and~\eqref{eq:RicciMI4D}, i.e.~$[\tilde{R}] = E^0$.}
\begin{equation}
  \label{eq:RicciMI}
R\ =\ \frac{4g_1}{g_0^2}\: +\: \left(\frac{2 g_1 g_0^{\prime} g_1^{\prime}}{g_0^3 k}-\frac{2
   g_1^2 g_0^{\prime\:2}}{g_0^4 k}-\frac{g_1^{\prime\:2}}{2 g_0^2 k}\right)(\overline{\psi}\psi)^2\;.
\end{equation}
Observe that $R$ is a Lorentz scalar, but not a real-valued expression due to the appearance of the fermionic bilinear term $(\overline{\psi}\psi)^2$. For $g_0=g_1=1$, the super-Ricci scalar simplifies to
\begin{equation}\label{eq:RicciMI0}
 R\ =\ 4\;.
\end{equation}
It is important to remark here that the same result~\eqref{eq:RicciMI0} would have been obtained in the absence of the bosonic field~$\phi$. Consequently, the non-vanishing field-space curvature arises from the non-linear terms in the fermion fields in $\zeta_\alpha$ through the model function $g_1$ in~\eqref{eq:zetaMI}.

The above consideration can be easily extended to a 4D version of the SG-QFT Model~I considered in~\eqref{eq:ModelI}. In this case,
$\gamma^\mu$ stand for the usual 4D Dirac matrices, and the 
Dirac fermion has four components: $\psi^\tran = (\psi_1\,, \psi_2\,, \psi_3\,, \psi_4)$. The 4D SG-QFT model has $(1|8)$ dimensions giving rise to rather lengthy expressions for the super-Riemann tensor, which we will not present here. Instead, we give the field-space super-Ricci scalar, 
\begin{equation}
   \label{eq:RicciMI4D}
\begin{aligned}
R\ =&\ \frac{24 g_1}{g_0^2}\,-\, \frac{24g_1^2}{g_0^3}(\overline{\psi}\psi) \,+\,\left(\frac{2 g_1 g_0^\prime g_1^\prime}{g_0^3 k}-\frac{2 g_1^2 g_0^{\prime\:2}}{g_0^4 k}-\frac{g_1^{\prime\:2}}{2 g_0^2 k}-\frac{4 g_1^3}{g_0^4}\right) (\overline{\psi}\psi)^2\\
 &\ +\left(-\frac{16 g_1^2 g_0^\prime g_1^\prime}{g_0^4 k}+\frac{16 g_1^3 g_0^{\prime\:2}}{g_0^5 k}+\frac{4
   g_1 g_1^{\prime\:2}}{g_0^3 k}+\frac{40 g_1^4}{g_0^5}\right) (\overline{\psi}\psi)^3\\
&\ +\left(\frac{80 g_1^3 g_0^\prime g_1^\prime}{g_0^5 k}-\frac{80 g_1^4 g_0^{\prime\:2}}{g_0^6 k}-\frac{20
   g_1^2 g_1^{\prime\:2}}{g_0^4 k}+\frac{20 g_1^5}{g_0^6}\right) (\overline{\psi}\psi)^4\; .
\end{aligned}
\end{equation}
For $g_0=g_1=1$, the field-space Ricci scalar takes on the simpler form,
\begin{equation}
   \label{eq:RicciMI04D}
    R\ =\ 24\: -\: 24\,(\overline{\psi}\psi)\: -\: 4\,(\overline{\psi}\psi)^2\: +\: 40\,(\overline{\psi}\psi)^3\: +\: 20\,(\overline{\psi}\psi)^4\; .
\end{equation}
We note that \eqref{eq:RicciMI04D} becomes identical to the result one would obtain in a system with two fermions in 2D. This should be expected, since the number of degrees of freedom and the structure of the Lagrangian~\eqref{eq:ModelI} are exactly the same for the two cases. 

Let us now discuss an important feature of the geometric construction of Lagrangian~\eqref{eq:ModelI}, and SG-QFTs in general. Specifically, one  may  notice that under a naive non-linear reparameterisation of the fermion fields,
\begin{equation}
  \label{eq:Nlinfermion}
    \widetilde{\psi}\: =\: \psi \:\sqrt{1+\overline{\psi}\psi}\;,\qquad  \widetilde{\overline{\psi}}\: =\: \sqrt{1+\overline{\psi}\psi}
\:\overline{\psi} \;,
\end{equation}
one can turn a standard (canonical) Dirac Lagrangian,
\begin{equation}
   \label{eq:LDirac}
    \mathcal{L}_{\text{D}}\ =\ \frac{i}{2}\,\left[\overline{\psi}(\slashed{\partial}\psi)\: -\:(\slashed{\partial}\overline{\psi})\psi\right]\,,
\end{equation}
into the Lagrangian~\eqref{eq:ModelI}, in which $g_0 = g_1 = 1$ and all remaining model functions are set to zero, $k = Y = V = 0$.
This would seem to suggest that  a curved field-space theory can be obtained from a flat one by means of a non-linear reparameterisation
like~\eqref{eq:Nlinfermion}, and vice-versa. However, within our SG-QFT framework, such a transformation is not possible. 
More explicitly, in an SG-QFT, the standard Dirac Lagrangian must be recast into the covariant form,
\begin{equation}
   \label{eq:LDiraczeta}
\mathcal{L}_{\text{D}}\: =\: \frac{i}{2}\, \zeta_\alpha\,\slashed{\partial} \Psi^\alpha\; ,
\end{equation}
with $\Psi^\a = \{ \psi\,, \overline{\psi}^\tran\}$ and $\zeta_\a = 
\{ \overline{\psi}\,, \psi^\tran\}$. Any change of the fermionic field chart, $\Psi^a \to \widetilde{\Psi}^a$, must be done according to the transformations,   
\begin{equation}
   \label{eq:JacobianD}
    \partial_\mu \widetilde{\Psi}^\a\ =\ \partial_\mu \Psi^\b\: _\b J^\a\;,\qquad 
    \widetilde{\zeta}_\a\ =\ \zeta_\b\: ^\b (J^{-1})^{\st}_{\a}\;.
\end{equation}
However, the Jacobian transformations~\eqref{eq:JacobianD} do not alter the form of $\mathcal{L}_{\text{D}}$ in~\eqref{eq:LDiraczeta}. Therefore, 
different analytic forms of the model function 
$\zeta_{\alpha}^{\mu}$ give rise to distinct supergeometric constructions of Lagrangians, involving different supermetrics $\boldsymbol{G}$. The superdeterminant of the latter usually affect the path-integral measure and so the effective action beyond the classical approximation~\cite{Finn:2019aip,Finn:2020nvn}.

Finally, we should comment on the flavour covariance of an SG-QFT with many species of fermions.
Indeed, an equivalent class of Lagrangians can be consistently constructed through flavour field redefinitions, $\boldsymbol{\psi} \to \boldsymbol{\widetilde{\psi}} = \boldsymbol{U} \boldsymbol{\psi}$,
where  $\boldsymbol{U}$ is a unitary flavour-rotation matrix that may only depend on the scalar fields. The new supermetric in the flavour-transformed basis is derived from the usual rank-2 covariance relation,
\begin{equation} 
   \label{tilde_metric}
    _\a \widetilde{G}_\b =\:_\a (J^{-1})^\c\:_\c G_\delta \:^\delta (J^{-1})^\st_\b\; .
\end{equation}
The so-derived supermetric in~\eqref{tilde_metric} can be shown to be equivalent to the supermetric that would be obtained by extracting the new model functions from a flavour-transformed Lagrangian${}$, e.g.~$\mathcal{\widetilde{L}}$. 
This last property provides further support of the mathematical and physical consistency of the SG-QFT framework under study.

\subsection{Non-zero Fermionic Field-Space Curvature: Model II}

We now turn our attention to the second category of SG-QFTs, for which 
the model function $\zeta^\mu_\a$ cannot be written in the factorisable form of~\eqref{eq:zetagamma}. For brevity, we call this scenario Model~II, in order to
distinguish from Model~I discussed in the previous subsection.

To showcase the rich geometric structure of this new class of SG-QFTs, we ignore all scalar fields and only consider one Dirac fermion $\psi$ in~4D. A minimal SG-QFT Model II is described by the Lagrangian,
\begin{equation}
   \label{eq:MII}
    \mathcal{L}_\mathrm{II}\ =\ \frac{i}{2}\left[\overline{\psi}\gamma^{\mu}(\partial_{\mu}\psi)-(\partial_{\mu}\overline{\psi})\gamma^{\mu}\psi\right]\: +\: \frac{i}{2}\overline{\psi}\gamma^{\mu}\psi\left[\overline{\psi}(\partial_{\mu}\psi)-(\partial_{\mu}\overline{\psi})\psi\right]\;.
\end{equation}
As outlined in Section~\ref{sec:SG}, we employ~\eqref{eq:kzeta} to calculate the model function $\zeta^\mu_\a$,
\begin{equation}
    \zeta^{\mu}_\a\:=\Big\{0\:,\:\overline{\psi}\gamma^{\mu}+(\overline{\psi}\gamma^{\mu}\psi)\,\overline{\psi}\:,\:\psi^\tran\gamma^{\mu\:\tran}+(\overline{\psi}\gamma^{\mu}\psi)\,\psi^\tran\Big\}\;.
\end{equation}
To extract the covector $\zeta_\a$ from this latter expression, we make use of the projection method given in~\eqref{eq:projzeta}. Following the approach of~\cite{Finn:2020nvn}, we construct the anti-supersymmetric rank-2 field-space tensor $_\a\lambda_\b$, which in turn was used to determine the vielbeins $_\a e^a$.
With the help of $_\a e^a$, the following field-space supermetric is derived:
\begin{equation}\label{eq:metricII}
    \boldsymbol{G}\: \equiv\: \{ _\a G_\b\}\: =\: \left(\begin{array}{rr}
 0 &\ d^\tran\\
 -d & 0\hspace{2mm}{}
\end{array}\right),
\end{equation}
where $d$ is a $4\times 4$-dimensional matrix in the Dirac spinor space given by
\begin{equation}
  \label{eq:MIId}
\begin{aligned}
d\: =\: 1_4\, +\, \frac{1}{4}(\overline{\psi}{\gamma}^{\mu}\psi)\,\gamma_\mu\, +\, \frac{1}{4}\gamma^{\mu}\,\psi\overline{\psi}\,\gamma_\mu\; .
    \end{aligned}
\end{equation}
Knowing the analytic form of the supermetric~$\boldsymbol{G}$, we may now compute the super-Ricci scalar of this theory, 
\begin{equation}
   \label{eq:MIIRscalar}
    R\: =\: -8+2 (\overline{\psi}\psi) +\frac{23}{8}(\overline{\psi}\psi)^2+\frac{9}{8}(\overline{\psi}\gamma_{5}\psi)^2 +\frac{5}{4}(\overline{\psi}\gamma_{\mu}\psi)(\overline{\psi}\gamma^{\mu}\psi) -\frac{29}{12}(\overline{\psi}\psi)^3 +\frac{7}{16}(\overline{\psi}\psi)^4\;.
\end{equation}
Observe that the super-Ricci scalar $R$ is both parity-preserving and Lorentz invariant, so it shares the same properties like the original Lagrangian~\eqref{eq:MII} from which it was obtained. Interestingly enough, the expression for $R$ of Model II has a much richer expansion than that found in Model~I [cf.~\eqref{eq:RicciMI04D}]. In addition to $(\overline{\psi}\psi)^2$ terms, $R$ now contains new Lorentz-invariant four-fermion operators, such as $(\overline{\psi}\gamma_{5}\psi)^2$ and  $(\overline{\psi}\gamma_{\mu}\psi)(\overline{\psi}\gamma^{\mu}\psi)$.

We should remark here that had we used the method of~\cite{Finn:2020nvn} given in~\eqref{eq:zeta} to deduce the covector $\zeta_\a$, we would then have obtained a different supermetric leading to an expression for $R$ similar to Model I as powers of the fermionic bilinears $\overline{\psi}\psi$ but with different coefficients. Nevertheless, we find that the projection method introduced in~\eqref{eq:projzeta} is more appropriate, since it reflects more accurately the geometric structure of the SG-QFT models in the second category. Furthermore, it should be noted that as opposed to Model I, the SG-QFT Model II cannot be brought into the canonical form of~\eqref{eq:LDirac} by naive redefinitions of the fermion fields: $\widetilde{\psi} = \psi\, f(\overline{\psi}\psi )$ and $\widetilde{\overline{\psi}} =  f^*(\overline{\psi}\psi )\,\overline{\psi}$, where $f$ is some judicious function like~\eqref{eq:Nlinfermion}. 

As we will see in the next section, field-space geometry governs the Feynman rules of an SG-QFT through superpropagators and supervertices.

\section{Superpropagators and Supervertices}\label{sec:Superprops}

In this section, we present analytical results of the superpropagator and 
the three- and four-point supervertices related to the fermionic part of SG-QFTs,
where the model function $_\a k_\b$ was set to zero.
However, we allow for the possible presence of background scalar fields~$\phi^A$.

In this simplified SG-QFT setting, we first give the equation of motion of the fields, 
\begin{equation}
  \label{eq:Sa}
S_{;\hat{\a}}\: =\: S_{,\hat{\a}}\: =\: i(-1)^{\a}\:_{\a}\lambda^{\mu}_{\rho}\, \partial_{\mu}\Phi^{\rho}\, -\, U_{,\a}\;.
\end{equation}
Here and in the following, a semicolon (;) stands for covariant configuration-space differentiation and  
$S$ is the classical action pertinent to the Lagrangian~\eqref{eq:LSGQFT}, with 
$_\a k_\b = 0$. In addition, we introduced in~\eqref{eq:Sa} a modified  version of the tensor ${}_{\alpha} \lambda_{\beta}$ of \eqref{eq:lambda} defined as
\begin{equation}
  \label{eq:lambdamu}
   {}_{\alpha} \lambda^\mu_{\beta}\ \equiv\ \frac{1}{2}\, \Big({}_{\a,}\zeta^\mu_{\beta}\: -\: (-1)^{\alpha+\beta+\alpha \beta}  \:_{\b,}\zeta^\mu_{\alpha}\Big)\; ,
\end{equation} 
which will appear in our expressions for the superpropagator and the supervertices given below. Notice that ${}_{\alpha} \lambda^\mu_{\beta}$ is a proper spacetime vector and rank-2 field-space tensor as it is derived from functional differentiation of $\zeta^\mu_{\alpha}$.

From~\eqref{eq:Sa}, the covariant inverse superpropagator $S_{;\hat{\a}\hat{\b}}$ may be evaluated
as follows:
\begin{equation}
  \label{eq:Sab}
S_{;\hat{\a}\hat{\b}}\: =\: i(-1)^{\a}\Big(
{}_\a\lambda^{\mu}_\rho\, \partial_{\mu}\Phi^\rho_{\ ;\b}\, +\,(-1)^{\rho\b} {}_\a\lambda^{\mu}_{\rho;\b}\,\partial_{\mu}\Phi^\rho\Big)\, -\, U_{;\a\b}\;.
\end{equation}
In the spacetime homogeneous limit of the theory in which $\partial_\mu \Phi \to 0$,
expression~\eqref{eq:Sab} becomes in momentum space,
\begin{equation}
   \label{eq:Sab0}
S_{;\hat{\a}\hat{\b}}\big|_{\partial_{\mu}\Phi=0}\, =\: \Big((-1)^{\a}\:_\a\lambda^{\mu}_\b\:p^\b_{\mu}\: -\: U_{;\a\b}\Big)\,\delta(p^\a+p^\b)\;.    
\end{equation}

Our next step is to calculate the covariant three-supervertex~$S_{;\hat{\a}\hat{\b}\hat{\c}}$. As before, we start evaluating this in the coordinate space, 
\begin{equation}
   \label{eq:Sabc}
\begin{aligned}
S_{;\hat{\a}\hat{\b}\hat{\c}}=i&(-1)^{\a}\Big( {}_\a\lambda^{\mu}_\rho\,\partial_{\mu}\Phi^\rho_{\ ;\b\c}
\, +\, (-1)^{\c(\rho+\b)}{}_\a\lambda^{\mu}_{\rho;\c}\,\partial_{\mu}\Phi^\rho_{\ ;\b}\\
    &\:+(-1)^{\rho\b} {}_\a\lambda^{\mu}_{\rho;\b}\,\partial_{\mu}\Phi^\rho_{\ ;\c}\,+\,(-1)^{\rho(\b+\c)}{}_\a\lambda^{\mu}_{\rho;\b\c}\partial_{\mu}\Phi^\rho\Big)\, -\, U_{;\a\b\c}\;.
\end{aligned}
\end{equation}
In the momentum space and homogeneous limit $\partial_\mu \Phi \to 0$ of the theory, the covariant three-supervertex reads
\begin{equation}
  \label{eq:Sabc0}
S_{;\hat{\a}\hat{\b}\hat{\c}}\big|_{\partial_{\mu}\Phi=0}\, =\: \Big(
    (-1)^{\a} {}_{\a}\lambda^{\mu}_{\b;\c}\,p^\b_{\mu}\, +\, (-1)^{\a+\b\c}{}_\a\lambda^{\mu}_{\c;\b}\,p^\c_{\mu}\:-\: U_{;\a\b\c}\Big)\,\delta (p^\a+p^\b+p^\c)\;.
\end{equation}
Notice that unlike a pure bosonic theory, the covariant three-supervertex $S_{;\hat{\a}\hat{\b}\hat{\c}}$ does not vanish in fermionic SG-QFTs in the absence of a potential term~$U$.

In a similar fashion, we can compute the four-supervertex in the configuration space as
\begin{equation}
\begin{aligned}
     S_{;\hat{\a}\hat{\b}\hat{\c}\hat{\d}}\: =\: &i(-1)^{\a}\Big( {}_\a\lambda^{\mu}_\rho\,\partial_{\mu}\Phi^\rho_{\ ;\b\c\delta}\, +\,(-1)^{\delta(\b+\c+\rho)}{}_\a\lambda^{\mu}_{\rho;\delta}\,\partial_{\mu}\Phi^\rho_{\ ;\b\c}\\
     &+(-1)^{\c(\b+\rho)}{}_\a\lambda^{\mu}_{\rho;\c}\,\partial_{\mu}\Phi^\rho_{\ ;\b\delta}\,+(-1)^{\rho\b}{}_\a\lambda^{\mu}_{\rho;\b}\,\partial_{\mu}\Phi^\rho_{\ ;\c\delta}\\
     &+\,(-1)^{(\c+\delta)(\rho+\b)}\,{}_\a\lambda^{\mu}_{\rho;\c\delta}\,\partial_{\mu}\Phi^\rho_{\ ;\b}\,+\,(-1)^{\rho(\b+\delta)+\delta\c}\, {}_\a\lambda^{\mu}_{\rho;\b\delta}\,\partial_{\mu}\Phi^\rho_{\ ;\c}\\
     &+(-1)^{\rho(\b+\c)}{}_\a\lambda^{\mu}_{\rho;\b\c}\,\partial_{\mu}\Phi^\rho_{\ ;\delta}\, +\,(-1)^{\rho(\b+\c+\delta)}\,{}_\a\lambda^{\mu}_{\rho;\b\c\delta}\,\partial_{\mu}\Phi^\rho\Big)\: -\: U_{;\a\b\c\delta}\; .
\end{aligned}
\end{equation}
In the momentum space and the homogeneous limit, the latter expression simplifies to
\begin{equation}
  \label{eq:Sabcd0}
\begin{aligned}
     \left. S_{;\hat{\a}\hat{\b}\hat{\c}\hat{\d}}\right|_{\partial_{\mu}\Phi=0}=&\ 
     \bigg( (-1)^{\a} {}_\a\lambda^{\mu}_\rho\: R^{\rho}_{\:\:\b\c\delta}\:p^\delta_{\mu}\:+\:(-1)^{\a} {}_\a\lambda^{\mu}_{\b;\c\delta}\:p^\b_{\mu}\:+\:(-1)^{\a+\b\c}{}_\a\lambda^{\mu}_{\c;\b\delta}\:p^\c_{\mu}\\
     &\ +\:(-1)^{\a+\delta(\b+\c)}{}_\a\lambda^{\mu}_{\delta;\b\c}\:p^\delta_{\mu}\ -\
U_{;\a\b\c\delta}\bigg)\,\delta (p^\a + p^\b +p^\c + p^\delta)\;.
\end{aligned}
\end{equation}

One can now make explicit the field-space metric dependence on the supervertices by writing ${}_\a\lambda^{\mu}_\b$ as: ${}_\a\lambda^{\mu}_\b = {}_\a(\lambda^\mu)^\rho\:{}_\rho G_\b$. 
When computing covariant derivatives of ${}_\a\lambda^{\mu}_\b$ through the contraction ${}_\a(\lambda^\mu)^\rho\:{}_\rho G_\b$, there will be a vanishing contribution arising from covariant derivatives of the supermetric $_\a G_\b$. One should bear in mind that $_\a G_\b$ satisfies the metric compatibility condition: 
\begin{equation}
    {}_\a G_{\b;\c}\:\equiv\: {}_\a G_{\b,\c}\:-\:{}_\a G_{\rho}\,\Gamma^{\rho}_{\:\:\b\c}\:-\:(-1)^{\a+\rho+\b(\rho+\a)}\,{}_\rho G_{\b}\,\Gamma^{\rho}_{\:\:\a\c}\ =\ 0\; ,
\end{equation}
and so non-zero contributions can only come from covariant differentiations of ${}_\a(\lambda^\mu)^\b$. Hence, it is this misalignment between ${}_\a\lambda^{\mu}_\b$ and ${}_\a G_\b$ that yields a non-vanishing three-supervertex $S_{;\hat{\a}\hat{\b}\hat{\c}}$ in~\eqref{eq:Sabc0}, even in the absence of potential terms. This is in contrast to the bosonic case as shown  in~\cite{Finn:2019aip,Cohen:2021ucp}.

We may now verify that the four-supervertices given in \eqref{eq:Sabcd0} satisfy two essential super-Ricci identities involving the \textit{supercommutator} of covariant derivatives.
First, we remind the reader that the supercommutator of two covariant derivatives is defined as~\cite{DeWitt:2012mdz}:
\begin{equation}
S_{;[\hat{\alpha},\hat{\beta}]} \:\equiv\:S\left[\overleftarrow{\nabla}_{\hat{\alpha}}\,,\overleftarrow{\nabla}_{\hat{\beta}}\right]\:=\:  S_{;\hat{\alpha}\hat{\beta}}\:-\:(-1)^{\hat{\alpha}\hat{\beta}}\,S_{;\hat{\beta}\hat{\alpha}}\;.
\end{equation}
Then, one can show that the following super-Ricci identities are satisfied in the homogeneous limit:
\begin{equation}
    \begin{aligned}      \left.S_{;\hat{\alpha}[\hat{\beta},\hat{\gamma}]\hat{\delta}}\right|_{\partial_{\mu}\Phi=0}&\:=\:(-1)^{\hat{\delta}(\hat{\rho}+\hat{\alpha}+\hat{\beta}+\hat{\gamma})} \,\left.S_{;\hat{\rho}\hat{\delta}}\right|_{\partial_{\mu}\Phi=0}\,R^{\hat{\rho}}_{\:\:\hat{\alpha} \hat{\beta}\hat{\gamma}}\;,\\    \left.S_{;\hat{\alpha}\hat{\beta}[\hat{\gamma},\hat{\delta}]}\right|_{\partial_{\mu}\Phi=0}&\:=\: \left.S_{;\hat{\alpha} \hat{\rho}}\right|_{\partial_{\mu}\Phi=0}\,R^{\hat{\rho}}_{\:\:\hat{\beta}\hat{\gamma}\hat{\delta}}\:+\:(-1)^{\hat{\beta}(\hat{\rho}+\hat{\alpha})}\, \left.S_{;\hat{\rho}\hat{\beta}}\right|_{\partial_{\mu}\Phi=0}\,R^{\hat{\rho}}_{\:\: \hat{\alpha}\hat{\gamma}\hat{\delta}}\; .
    \end{aligned}
\end{equation}
These identities turn out to be rather useful in simplifying the process of supersymmetrisation of higher-point supervertices.

We conclude this section by noting that pure scalar 
contributions to the superpropagators and supervertices~\cite{Honerkamp:1971sh, Ecker:1972tii, Alonso:2016oah}
can also be included in the above expressions. Unlike the
fermionic contributions which depend linearly on particle momenta, bosonic effects are quadratic in the momenta and so they enter additively to~\eqref{eq:Sab0}, \eqref{eq:Sabc0} and \eqref{eq:Sabcd0}.

\section{Summary and Outlook}\label{sec:Summary}

We have studied in detail the frame-covariant formalism presented earlier in~\cite{Finn:2020nvn} on scalar-fermion theories. The scalar and fermion fields define a coordinate system or a chart which describe a supermanifold in the configuration space of the respective QFTs.  We discussed the issue of uniqueness of the supermetric and clarified that different choices of the latter lead to distinct Supergeometric QFTs in the off-shell kinematic region, as well as beyond the classical approximation. 

Adopting a natural and self-consistent choice for the supermetric, we have shown that scalar fields alone do not provide a new source of curvature in the fermionic sector of the theory beyond the one that originates from the model function ${}_\alpha k_\beta$. In particular, we have explicitly demonstrated that non-linear powers of fermionic fields in the model function $\zeta^\mu_\alpha$ can give rise
to non-zero fermionic curvature, as expressed by a non-zero super-Riemann tensor. Hence, we have 
presented for the first time novel minimal SG-QFT models that feature non-zero fermionic curvature both in two and four spacetime dimensions up to second order in spacetime derivatives. It~should be emphasised here that the resulting super-Riemann tensor and super-Ricci scalar may contain fermionic bilinears which are no proper real numbers. This should be contrasted with Supergravity theories~\cite{Alvarez-Gaume:1981exa} where the curvature is a real-valued expression dictated by the scalar part of the Kaehler manifold, on which the fermions were treated as tangent vectors. 

In addition, we have derived new generalised expressions for the scalar-fermion inverse super\-propagator, and the three- and four-supervertices. As opposed to pure bosonic theories, we have found that the three-supervertices are non-zero in fermionic theories in the absence of a zero-grading scalar potential~$U$~[cf.~\eqref{eq:LSGQFT}]. These ingredients are all necessary for future considerations in evaluating amplitudes and higher-loop effective actions in SG-QFTs. Furthermore, one may wish to include further gauge and gravitational symmetries in SG-QFTs which will act as isometries~\cite{Vilkovisky:1984st,DeWitt:1985sg} on the supermanifold. We expect that SG-QFTs will lead to
a complete geometrisation of realistic theories of micro-cosmos, such as the SM and its gravitational sector. We may even envisage that SG-QFTs will provide a new portal to the dark sector, where dark-sector fermionic fields may modify the dispersion properties of weakly interacting particles, like SM neutrinos and axions.
We plan to investigate the above issues in future works.

\subsection*{Acknowledgements} 
The authors thank Alejo Rossia and Thomas McKelvey for discussions. 
The work of AP is supported in part by the STFC Research Grant ST/T001038/1. VG acknowledges support by the University of Manchester through the President's Doctoral Scholar Award. 

\newpage

\bibliography{SG_QFT}

\providecommand{\href}[2]{#2}\begingroup\raggedright\begin{thebibliography}{10}

\bibitem{DeWitt:1967ub}
B.S.~DeWitt, \emph{{Quantum Theory of Gravity. 2. The Manifestly Covariant
  Theory}}, \href{https://doi.org/10.1103/PhysRev.162.1195}{\emph{Phys. Rev.}
  {\bfseries 162} (1967) 1195}.

\bibitem{Gaillard:1985uh}
M.K.~Gaillard, \emph{{The Effective One Loop Lagrangian With Derivative
  Couplings}}, \href{https://doi.org/10.1016/0550-3213(86)90264-6}{\emph{Nucl.
  Phys. B} {\bfseries 268} (1986) 669}.

\bibitem{Pilaftsis:1996fh}
A.~Pilaftsis, \emph{{Generalized Pinch Technique and the Background Field
  Method in General Gauges}},
  \href{https://doi.org/10.1016/S0550-3213(96)00686-4}{\emph{Nucl. Phys. B}
  {\bfseries 487} (1997) 467}
  [\href{https://arxiv.org/abs/hep-ph/9607451}{{\ttfamily hep-ph/9607451}}].

\bibitem{Cornwall:2010upa}
J.M.~Cornwall, J.~Papavassiliou and D.~Binosi, \emph{{The Pinch Technique and
  its Applications to Non-Abelian Gauge Theories}}, Cambridge University Press
  (12, 2010).

\bibitem{Binosi:2009qm}
D.~Binosi and J.~Papavassiliou, \emph{{Pinch Technique: Theory and
  Applications}},
  \href{https://doi.org/10.1016/j.physrep.2009.05.001}{\emph{Phys. Rept.}
  {\bfseries 479} (2009) 1} [\href{https://arxiv.org/abs/0909.2536}{{\ttfamily
  0909.2536}}].

\bibitem{Honerkamp:1971sh}
J.~Honerkamp, \emph{{Chiral multiloops}},
  \href{https://doi.org/10.1016/0550-3213(72)90299-4}{\emph{Nucl. Phys. B}
  {\bfseries 36} (1972) 130}.

\bibitem{Ecker:1972tii}
G.~Ecker and J.~Honerkamp, \emph{{Covariant perturbation theory and chiral
  superpropagators}},
  \href{https://doi.org/10.1016/0370-2693(72)90074-3}{\emph{Phys. Lett. B}
  {\bfseries 42} (1972) 253}.

\bibitem{Alvarez-Gaume:1981exa}
L.~Alvarez-Gaume, D.Z.~Freedman and S.~Mukhi, \emph{{The Background Field
  Method and the Ultraviolet Structure of the Supersymmetric Nonlinear Sigma
  Model}}, \href{https://doi.org/10.1016/0003-4916(81)90006-3}{\emph{Annals
  Phys.} {\bfseries 134} (1981) 85}.

\bibitem{Vilkovisky:1984st}
G.A.~Vilkovisky, \emph{{The Unique Effective Action in Quantum Field Theory}},
  \href{https://doi.org/10.1016/0550-3213(84)90228-1}{\emph{Nucl. Phys. B}
  {\bfseries 234} (1984) 125}.

\bibitem{DeWitt:1985sg}
B.S.~DeWitt, \emph{{The Effective Action}},  in \emph{{Les Houches School of
  Theoretical Physics: Architecture of Fundamental Interactions at Short
  Distances: Proceedings, Les Houches 44th Summer School of Theoretical
  Physics: Les Houches, France, July 1-August 8, 1985}}, pp.~1023--1058.

\bibitem{Barvinsky:1985an}
A.O.~Barvinsky and G.A.~Vilkovisky, \emph{{The Generalized Schwinger-Dewitt
  Technique in Gauge Theories and Quantum Gravity}},
  \href{https://doi.org/10.1016/0370-1573(85)90148-6}{\emph{Phys. Rept.}
  {\bfseries 119} (1985) 1}.

\bibitem{Ellicott:1987ir}
P.~Ellicott and D.J.~Toms, \emph{{On the New Effective Action in Quantum Field
  Theory}}, \href{https://doi.org/10.1016/0550-3213(89)90579-8}{\emph{Nucl.
  Phys. B} {\bfseries 312} (1989) 700}.

\bibitem{Burgess:1987zi}
C.P.~Burgess and G.~Kunstatter, \emph{{On the Physical Interpretation of the
  Vilkovisky-de Witt Effective Action}},
  \href{https://doi.org/10.1142/S0217732387001117}{\emph{Mod. Phys. Lett. A}
  {\bfseries 2} (1987) 875}.

\bibitem{Odintsov:1989gz}
S.D.~Odintsov, \emph{{The Parametrization Invariant and Gauge Invariant
  Effective Actions in Quantum Field Theory}}, {\emph{Fortsch. Phys.}
  {\bfseries 38} (1990) 371}.

\bibitem{Kamenshchik:2014waa}
A.Y.~Kamenshchik and C.F.~Steinwachs, \emph{{Question of quantum equivalence
  between Jordan frame and Einstein frame}},
  \href{https://doi.org/10.1103/PhysRevD.91.084033}{\emph{Phys. Rev. D}
  {\bfseries 91} (2015) 084033}
  [\href{https://arxiv.org/abs/1408.5769}{{\ttfamily 1408.5769}}].

\bibitem{Burns:2016ric}
D.~Burns, S.~Karamitsos and A.~Pilaftsis, \emph{{Frame-Covariant Formulation of
  Inflation in Scalar-Curvature Theories}},
  \href{https://doi.org/10.1016/j.nuclphysb.2016.04.036}{\emph{Nucl. Phys. B}
  {\bfseries 907} (2016) 785}
  [\href{https://arxiv.org/abs/1603.03730}{{\ttfamily 1603.03730}}].

\bibitem{Karamitsos:2017elm}
S.~Karamitsos and A.~Pilaftsis, \emph{{Frame Covariant Nonminimal Multifield
  Inflation}},
  \href{https://doi.org/10.1016/j.nuclphysb.2017.12.015}{\emph{Nucl. Phys. B}
  {\bfseries 927} (2018) 219}
  [\href{https://arxiv.org/abs/1706.07011}{{\ttfamily 1706.07011}}].

\bibitem{Finn:2019aip}
K.~Finn, S.~Karamitsos and A.~Pilaftsis, \emph{{Frame Covariance in Quantum
  Gravity}}, \href{https://doi.org/10.1103/PhysRevD.102.045014}{\emph{Phys.
  Rev. D} {\bfseries 102} (2020) 045014}
  [\href{https://arxiv.org/abs/1910.06661}{{\ttfamily 1910.06661}}].

\bibitem{Falls:2018olk}
K.~Falls and M.~Herrero-Valea, \emph{{Frame (In)equivalence in Quantum Field
  Theory and Cosmology}},
  \href{https://doi.org/10.1140/epjc/s10052-019-7070-3}{\emph{Eur. Phys. J. C}
  {\bfseries 79} (2019) 595}
  [\href{https://arxiv.org/abs/1812.08187}{{\ttfamily 1812.08187}}].

\bibitem{Alonso:2016oah}
R.~Alonso, E.E.~Jenkins and A.V.~Manohar, \emph{{Geometry of the Scalar
  Sector}}, \href{https://doi.org/10.1007/JHEP08(2016)101}{\emph{JHEP}
  {\bfseries 08} (2016) 101}
  [\href{https://arxiv.org/abs/1605.03602}{{\ttfamily 1605.03602}}].

\bibitem{Nagai:2019tgi}
R.~Nagai, M.~Tanabashi, K.~Tsumura and Y.~Uchida, \emph{{Symmetry and geometry
  in a generalized Higgs effective field theory: Finiteness of oblique
  corrections versus perturbative unitarity}},
  \href{https://doi.org/10.1103/PhysRevD.100.075020}{\emph{Phys. Rev. D}
  {\bfseries 100} (2019) 075020}
  [\href{https://arxiv.org/abs/1904.07618}{{\ttfamily 1904.07618}}].

\bibitem{Helset:2020yio}
A.~Helset, A.~Martin and M.~Trott, \emph{{The Geometric Standard Model
  Effective Field Theory}},
  \href{https://doi.org/10.1007/JHEP03(2020)163}{\emph{JHEP} {\bfseries 03}
  (2020) 163} [\href{https://arxiv.org/abs/2001.01453}{{\ttfamily
  2001.01453}}].

\bibitem{Cohen:2021ucp}
T.~Cohen, N.~Craig, X.~Lu and D.~Sutherland, \emph{{Unitarity violation and the
  geometry of Higgs EFTs}},
  \href{https://doi.org/10.1007/JHEP12(2021)003}{\emph{JHEP} {\bfseries 12}
  (2021) 003} [\href{https://arxiv.org/abs/2108.03240}{{\ttfamily
  2108.03240}}].

\bibitem{Talbert:2022unj}
J.~Talbert, \emph{{The geometric \ensuremath{\nu}SMEFT: operators and
  connections}}, \href{https://doi.org/10.1007/JHEP01(2023)069}{\emph{JHEP}
  {\bfseries 01} (2023) 069}
  [\href{https://arxiv.org/abs/2208.11139}{{\ttfamily 2208.11139}}].

\bibitem{Helset:2022tlf}
A.~Helset, E.E.~Jenkins and A.V.~Manohar, \emph{{Geometry in scattering
  amplitudes}}, \href{https://doi.org/10.1103/PhysRevD.106.116018}{\emph{Phys.
  Rev. D} {\bfseries 106} (2022) 116018}
  [\href{https://arxiv.org/abs/2210.08000}{{\ttfamily 2210.08000}}].

\bibitem{Isidori:2023pyp}
G.~Isidori, F.~Wilsch and D.~Wyler, \emph{{The Standard Model effective field
  theory at work}},  \href{https://arxiv.org/abs/2303.16922}{{\ttfamily
  2303.16922}}.

\bibitem{Fumagalli:2020ody}
J.~Fumagalli, M.~Postma and M.~Van Den~Bout, \emph{{Matching and running
  sensitivity in non-renormalizable inflationary models}},
  \href{https://doi.org/10.1007/JHEP09(2020)114}{\emph{JHEP} {\bfseries 09}
  (2020) 114} [\href{https://arxiv.org/abs/2005.05905}{{\ttfamily
  2005.05905}}].

\bibitem{DeWitt:2012mdz}
B.S.~DeWitt, \emph{{Supermanifolds}}, Cambridge Monographs on Mathematical
  Physics, Cambridge Univ. Press, Cambridge, UK (5, 2012),
  \href{https://doi.org/10.1017/CBO9780511564000}{10.1017/CBO9780511564000}.

\bibitem{Finn:2020nvn}
K.~Finn, S.~Karamitsos and A.~Pilaftsis, \emph{{Frame covariant formalism for
  fermionic theories}},
  \href{https://doi.org/10.1140/epjc/s10052-021-09360-w}{\emph{Eur. Phys. J. C}
  {\bfseries 81} (2021) 572}
  [\href{https://arxiv.org/abs/2006.05831}{{\ttfamily 2006.05831}}].

\bibitem{Pilaftsis:2022las}
A.~Pilaftsis, K.~Finn, V.~Gattus and S.~Karamitsos, \emph{{Geometrising the
  Micro-Cosmos on a Supermanifold}},
  \href{https://doi.org/10.22323/1.406.0080}{\emph{PoS} {\bfseries CORFU2021}
  (2022) 080} [\href{https://arxiv.org/abs/2204.00123}{{\ttfamily
  2204.00123}}].

\bibitem{Schwinger:1963re}
J.S.~Schwinger, \emph{{Quantized gravitational field}},
  \href{https://doi.org/10.1103/PhysRev.130.1253}{\emph{Phys. Rev.} {\bfseries
  130} (1963) 1253}.

\bibitem{Yepez:2011bw}
J.~Yepez, \emph{{Einstein's vierbein field theory of curved space}},
  \href{https://arxiv.org/abs/1106.2037}{{\ttfamily 1106.2037}}.

\end{thebibliography}\endgroup


\begin{thebibliography}{99}

\bibitem{DeWitt:1967ub}
B. S. DeWitt,
\emph{Quantum Theory of Gravity. 2. The Manifestly Covariant Theory},
\href{https://doi.org/10.1103/PhysRev.162.1195}
{\emph{Phys. Rev.} {\bf 162} (1967) pg. 1195--1239}


\bibitem{b}
Author,
\emph{Title},
arxiv:1234.5678.

\bibitem{c}
Author,
\emph{Title},
Publisher (year).

\end{thebibliography}

\end{document}